\documentclass[twocolumn]{aastex62}
\usepackage{graphicx}

\newcommand{\EQ}{\begin{equation}}
\newcommand{\EN}{\end{equation}}
\newcommand{\EQA}{\begin{eqnarray}}
\newcommand{\ENA}{\end{eqnarray}}
\newcommand{\yr}{\,{\rm yr}}
\newcommand{\au}{\,{\rm AU}}
\newcommand{\pc}{\,{\rm pc}}
\newcommand{\msol}{\,{\rm M_{\odot}}}
\newcommand{\cmn}{\,{\rm cm^{-3}}}
\newcommand{\kms}{\,{\rm km\,s^{-1}}}

\received{December16, 2019}
\accepted{August 9, 2020}

\shorttitle{Survival of Population III stars till present day}
\shortauthors{Dutta et al.}

\begin{document}

\title{Modeling the survival of Population III stars till present day}

\author[0000-0001-5254-0621]{Jayanta Dutta}
\email{{\tt jd.astrop@gmail.com}}
\affil{Department of Physical Sciences, Indian Institute of Science Education and Research (IISER),~Mohali,~140306,~India}
\author{Sharanya Sur}
\affil{Indian Institute of Astrophysics, 2nd Block, Koramangala, Bangalore 560034,~India}
\author{Athena Stacy}
\affil{University of California,~Berkeley,~CA 94720,~USA}
\author{Jasjeet Singh Bagla} 
\affil{Department of Physical Sciences, Indian Institute of Science Education and Research (IISER),~Mohali,~140306,~India}

%%%%%%%%%%%%%%%%%%%%%%%%%%%%%%%
\begin{abstract}

Recent numerical simulations have suggested the probability of a fraction of the primordial stars being 
ejected from the cluster of their origin. We explore the possibility that some of these can remain on the 
main sequence until the present epoch. We develop a semianalytical model guided by results of cosmological 
simulations to study the mass accretion by these protostars as a function of the original stellar mass and 
other parameters such as angular momentum and gravitational drag due to ambient gas. We also explore 
whether some of the protostars remain sufficiently low mass and long-lived to survive to the present day. 
This requires that the protostars are ejected from the star forming region while their mass is less than 
$0.8\msol$. Assuming that the protostars gain mass via the spherical Bondi--Hoyle accretion from the ambient 
medium, we show that Population III protostars that initially form within a certain range of mass and are ejected 
with velocity larger than the escape velocity may survive to the present day on the main sequence. Thus, 
they may even be found in our Milky Way or its satellites. Our calculations also reveal that protostars that 
do not get ejected from the parent gas clump accrete a large amount of gas. We predict that these can 
become massive enough to be progenitors of black holes.
\end{abstract}


\section{Introduction}
\label{sec:intro}

With the advent of state-of-the art observational technology and subsequent release of enormous data 
sets from modern telescope, satellite, and wide-range surveys, the study of emergence of the very first 
stars in the universe has become an important topic of research in modern astrophysics and cosmology 
\cite[see, e.g.][for recent observational predictions]{caffau11,frebel15,bonifacio18,Hashimoto18,hartwig19}.
Hierarchical structure formation leads to the formation of small gravitationally bound objects such as 
matter peaks containing the first stars, also known as Population III or Pop III stars 
\citep{sharda19,susa19,sugimura20,wollenberg20}. The entire process is highly nonlinear in 
nature and a natural outcome of the well-defined cold dark matter cosmology, aka $\Lambda$CDM 
model, where smaller structures subsequently merge with other halos to form larger objects like 
galaxies and clusters \citep{springel06,by11,haiman11,hirano14,greif15,wise19}.
Young stars in earliest galaxies emit ultraviolet (UV) radiation that ionizes the surrounding intergalactic 
medium, thereby terminating the cosmic dark ages \citep{gao07,bkp09,kbb13,sluder16}. 

In spite of tremendous advances in our understanding of how the first stars form, the complex nature of 
the nonlinear process during the gravitational collapse of primordial gas renders many important details 
associated with their formation uncertain \citep{xu16,barrow17}.
For example, results from early simulations of collapse of the unstable gas clumps suggested that these 
stars may have very large masses of around $100\,M_\odot$ \citep{abn02,on07,yoh08,woods17}. 
However, this conclusion was likely due to computational limitation that prevented following the evolution 
of gas physics for adequately long periods of time at significantly high resolution. In particular, the 
requirement of a time step of $\sim 10^{-3}\yr$ \citep[e.g.,][]{abn02,bl03} made it impossible to follow the 
collapse to sufficiently high densities. With the development of sophisticated numerical algorithms and 
techniques, subsequent 3D simulations circumvented the above limitation by introducing 
sink particles above a certain density threshold \cite[e.g.,][]{krumholz04,cgkb11,dnck15,sbl16,sormani17}. 
This opened up a new window of following the evolution of unstable gas clumps beyond the formation of 
the first protostar, raising the possibility of probing the ultimate fate of the parent clumps inside dark matter 
halos. Indeed, these simulations showed that eventual fragmentation of the unstable circumstellar disk led 
to the formation of multiple protostars that are much smaller in mass \cite[e.g.,][]{cgsgkb11,greif12,md13}.

Solution of the aforementioned issue brought to the fore another interesting question. What is the ultimate 
fate of these evolving fragments that have a comparatively low mass, and which eventually become 
protostars after interaction with the surrounding gas? Two possible scenarios that emerge are the following: 
the fragments can migrate on the viscous time scale over which the angular momentum is lost during the 
entire evolution process, and hence move toward the center, eventually merging with the primary protostar 
on a scale of $\sim 10^4\au$ \cite[e.g.,][]{hb17}. Alternatively, the secondary protostars can escape the 
potential well of the bound system owing to gravitational interactions with each other and with the surrounding 
medium \citep{ishiyama16}. In an earlier study \citep{dutta16a}, it was shown that a fraction of the original 
protostars can be ejected from the cluster of their origin with speeds of $\sim 10 - 20\kms$, comparable to or 
larger than the escape speed of the system. If this happens, it is plausible that some of these protostars could 
enter the main sequence and might have survived until the present epoch, provided that they were unable to 
accrete significant mass before being ejected as low-mass stars with masses $\leq 0.8\msol$ \citep{marigo+01,
komiya09,komiya15,kti19,susa19}. This possibility of their survival has been answered somewhat positively in 
a few studies \citep{bond81,nu01}. However, a more careful investigation is required to understand the history 
of evolution processes, instabilities, interactions due to gravitational drag, trajectories of stars, and the mass 
accretion phenomenon, respectively. An insight into these physical processes is therefore crucial to study the 
survival of Population III stars for billions of years as the universe continues to evolve to the present complicated state. 

However, despite rapid progress in advanced numerical techniques with sophisticated Lagrangian and 
Eulerian codes, there are still many issues that need to be resolved. First, numerically simulating the 
complex interplay between the dynamical interaction of the fragments with the ambient gas and the accretion 
phenomena, as well as feedback, needs to be incorporated and tested at the relevant scales. Second, the 
computation of nonlinear gravitational gas collapse that follows the evolution beyond the formation of the 
first protostar suffers from shortcomings such as unphysical fragmentation and artificial viscosity, which 
basically decides the amount of kinetic energy that is thermalized in smoothed particle hydrodynamics 
\cite[SPH;][]{monaghan92,springel10sph}. The study by \citet{yoha06} has suggested that the artificial 
viscosity in Lagrangian hydrodynamics has minimal effects on the angular momentum transport during the
collapse of primordial gas. Moreover, highest-resolution 3D numerical simulations available at present are 
only able to follow about a thousand years of evolution after the formation of the first protostar \citep{bgsh15}. 
Thus, integration over a realistic number of orbital revolutions is well beyond the existing numerical capabilities. 

In view of the above-mentioned challenges, it is therefore important to explore whether the fate of the primordial 
protostars can be addressed by using simple semianalytical models, so as to have a preliminary 
understanding of the dynamics before delving into more complicated details using 3D simulation techniques. 
This is the approach that we adopt here. Specifically, we begin by using the typical orbital parameters of 
these protostars and the properties of the clusters such as density, temperature, and velocity, as determined 
from 3D numerical simulations of primordial star-forming gas clumps. This approach is advantageous in the 
sense that it allows us to use a range of parameters required to investigate the properties of the gas evolution 
as inputs into a semianalytical model with Bondi--Hoyle flow \citep{bondi44,bondi52} that accounts for the 
interaction of the protostars with the ambient gas. As detailed in the subsequent sections, this particularly 
enables us to probe the details of the trajectory of these stars that escape while interacting with the ambient 
gas in the gravitationally bound system. Further, it also allows us to explore the range of masses of 
Population III protostars that could have avoided core collapse and survived to the present day.

The paper is organized in the following manner. In section~\ref{sec:simulation} we describe in detail the 
numerical setup, which includes a comprehensive discussion of the initial conditions and an adaptive 
time-stepping scheme for the numerical integration. The details of dynamics such as the trajectories, mass 
accretion rates, and mass-velocity relations are outlined in section~\ref{sec:result}, followed by a summary of the main results and the implication of this study for the possible existence of Population III stars until the present epoch in section~\ref{sec:summary}. 

\section{Numerical Setup}
\label{sec:simulation}

The numerical technique should enable us to investigate in detail the long-term evolution of a system with 
multiple fragments. To this effect, we describe below the initial conditions of these protostars followed by 
our implementation of a simple model of spherical accretion, namely, Bondi--Hoyle (BH) accretion and 
dynamical evolution. 

\subsection{Initial Conditions}
\label{subsec:initial}

Theoretical calculations along with 1D and 3D numerical simulations have shown that the density 
profile of the primordial gas clumps in which first stars are formed is a well-defined power law of the 
form $\rho \propto r^{-2.2}$, corresponding to the typical distribution for a similar solution with 
$\gamma_{\rm eff} \sim 1.09$ \citep{suto88,on98}. We therefore model an unstable collapsing gas 
clump with a central core of size $r_0$ in this clump, such that the density profile is given by 
\EQ
\rho(r) \approx \frac{\rho_0}{(1 + r/r_0)^{2.2}} \,,
\label{denprof}
\EN
where the number density of gas is usually defined as $n(r) \sim \rho(r)/\mu\,m_p$, with an approximate 
mean molecular weight of the gas particle $\mu \approx 2.33$ and $m_p$ being the proton mass. 
Since our main interest here is on the mass accretion phase, we take the central number 
density of the clumps in the range $n_{0} \sim 10^{12} - 10^{13}\cmn$.

In a majority of runs whose results we report here, the core radius is taken to be $r_{0} \sim 5\au$. 
However, because the central regime is expected to evolve with time, we have additionally explored in 
Section~\ref{varnoro} the effect of different choices of $n_{0}$ and $r_{0}$ on the dynamical 
evolution of the protostars.
Considering that $m_p$ is the mass of a proton and $k_{\rm B}$ is the Boltzmann constant, the 
gas temperature and the sound speed are then estimated to be of the following form, with the 
core temperature approaching close to 1200 K \cite[again consistent with simulations; e.g.,][]{tao09,sur10}:
\EQ
T (r) \approx 1200 {\rm K} \times \left[\frac{\rho(r)}{\rho_0}\right]^{0.1}\,,\nonumber
\label{temprof}
\EN
\EQ
c_s (r) \approx \left[\frac{5\,k_B\,T(r)}{3 \times 2.33\,m_p}\right]^{0.5} \,.
\label{csprof}
\EN

Given the physical properties of the ambient medium, we now need to explore a set of initial conditions that 
covers the entire range of the parameter space. Because the fragmentation of the circumstellar disk occurs 
at various scales, the time of fragmentation depends strongly on the initial configuration of the clumps such 
as the degree of rotation, chemical abundances, hydrogen formation, and associated cooling and heating 
processes \citep{greif12,dutta15,dutta16b}. We therefore directly use the parameters such as radial and 
rotational velocities, acceleration, initial mass of evolving fragments, temperature, and density profile 
from the above-mentioned avant-garde cosmological simulations that represent the physical properties of this 
dynamical system of multiple fragments. For our numerical setup, we denote these fragments as primordial 
protostars that keep on evolving while moving around in this multiple system, which can be assumed to be a
small cluster of evolving protostars. We thus have the initial configuration to cover a large parameter space for 
studying the evolving system that consists of the following:

\begin{itemize}
\item
The protostars formed out of the slowly rotating gas clumps are placed near the center, whereas the others 
spread over larger radii owing to conservation of angular momentum. The evolving protostars are thus scattered 
at different positions that follow a power-law relation with the clump's initial degree of rotational support 
\citep{dutta16a}. In our semianalytical calculations, the initial positions of protostars ($i$) are distributed within 
the clumps in the range of $\sim 1 - 100\au$ from the central core of the cluster \citep{stacy13,greif15}.

\item
Protostars are aligned at an initial azimuthal angle $\phi \approx 0$. They are assigned radial and azimuthal 
components of initial velocity, $ v_{r,i}$ and $v_{\phi,i}$, respectively, and the corresponding acceleration 
$a_{r,i}$ and $a_{\phi,i}$. At the time of disk fragmentation, the newly formed fragments can have mass as 
low as $\sim 10^{-3}\msol$ \citep{abn02,yoh08}. For our analysis, we have chosen the evolving protostars 
with initial masses in the range of $0.01 \le M_{*,i} \le 1\msol$, which move with radial velocities 
$0 \le v_{r,i} \le 30\kms$ and azimuthal velocities $0.01 \le v_{\phi,i} \le 5\kms$, respectively. 

\item
Protostars are considered to be ejected out of the system if and when they reach a radial distance 
$r \sim 2\pc$, which is the typical size of the clumps \cite[e.g.,][]{by11}. Note that this scale is more than 
four orders of magnitude larger than the core radius. The total speed of the protostars relative to the 
ambient gas is, $v = (v_{r}^{2} + v_{\phi}^{2})^{0.5} \gg c_{s}$, where $c_s$ is the speed of sound of the 
ambient gas. 
\end{itemize}

We then use this initial configuration of the set of parameters for post-processing and compute the escape 
velocity, which is a function of distance from the central core. The escape speed corresponding to the density 
profile $\rho(r) \sim \rho_0 \, (1 + r/r_0)^{-2}$ in the ambient medium is given by 
\EQA
v_{esc}&&=\sqrt{{2GM_{\rm enc}(r_i) \over r_i}}\nonumber\\
&& =\Bigl [ (8 \pi G \rho_0 r_0^2) \Bigl ( {(1+r_i/r_o) \over r_i/r_0}
-{1 \over (r_i/r_0) (1+r_i/r_0)} \nonumber\\
&& \quad -{2 \ln (1+r_i/r_0) \over (r_i/r_0)} \Bigr ) \Bigr ]^{1/2} \,. 
\label{escape}
\ENA

Here $G$ is the gravitational constant and $M_{\rm enc}(r_i)$ represents the gas mass enclosed by the 
protostars that are placed at different distances $r_i$. This analytical expression provides an approximate 
estimate of the escape velocity that is close to values corresponding to the actual density profile in 
Equation (\ref{denprof}). As discussed, in general, protostars born out of more rapidly rotating gas clumps are 
situated away from the center to conserve angular momentum, whereas others are located around the 
center within a few au to tens of au. For Population III stars, the initial core radius is roughly a few au. Thus,
for a core radius $r_0 \approx 5\au$, the maximum escape speed is expected to be of order $\sim 11 - 13\kms$.  

\subsection{Equation of motion with Bondi--Hoyle flow}

Here we develop the basic numerical model that describes the dynamical evolution of the entire system. 
We recall from Section~\ref{sec:intro} that recent numerical simulations show that the unstable self-gravitating 
disk is prone to fragmentation, eventually leading to the formation of multiple protostars that, together with 
the ambient gas, are trapped in the gravitational potential well. This system can be formally described as a 
supersonic, compressible flow coupled to multiple gravitating, accreting, and potentially radiating bodies. 
The protostars, while moving through this medium, will experience a drag originating from the dynamical 
friction associated with star--cloud interactions. This drag force can cause a change in orbit akin to the 
case for X-ray binaries \citep{bdc17}. This can be approximated using Bondi--Hoyle accretion. 

Note that in its original form Bondi--Hoyle accretion considers the evolution of a mass moving through a uniform 
gas clump where it accretes material from the surrounding medium. We therefore write down the governing 
equations for the dynamics of the protostars traveling in the background of the primordial density distribution. 
As we shall see in the subsequent sections, this turns out to be an excellent approximation to describe the 
evolution of protostars orbiting in the disk while simultaneously accreting material from the disk.  

To start with, we consider protostars with initial mass $M_{*,i}$ that are orbiting with initial azimuthal velocity 
$v_{\phi, i}$, along with a range of radial velocity $v_{r, i}$. Here the subscript `$i$' stands for the number of 
protostars. We use the Bondi--Hoyle accretion rate to determine the time evolution of the stellar mass as 
\EQA
\frac{dM_*}{dt} = \frac{4\pi G^2M_*^2\rho(r)}{[c_s^2 + (v_{r}^2 + v_{\phi}^2)]^{3/2}} \,, 
\label{mass_acc}    
\ENA
where $\rho(r)$ corresponds to the density profile given by Equation~(1) and $c_s$ is the sound speed of the 
ambient medium.
\noindent
It is important to note that the Bondi--Hoyle flow cannot persist for long because protostars are increasing 
its mass and accumulating momentum. Therefore, a `dynamical friction' arises from the gravitational 
focusing of the ambient gas behind the protostar \citep{edgar04}. We therefore solve for the dynamics of 
the individual protostars (Equation~(\ref{motion})) along with conservation of angular momentum 
(Equation~(\ref{angmom})) and the mass accretion rate (Equation~(\ref{mass_acc})) using
\EQA
\frac{d^2r}{dt^2} = -\frac{GM_{\rm enc}(r)}{r^2} + \frac{v_{\rm
    \phi}^2}{r} - \frac{Av_r}{M_*{[c_s^2 + (v_{r}^2 + v_{\phi}^2)]}^{3/2}}\,, 
    \label{motion}
\end{eqnarray}
\begin{eqnarray}
\frac{d^2\phi}{dt^2} = -\frac{2v_r\dot{\phi}}{r} -
\frac{A\dot{\phi}}{M_*{[c_s^2 + (v_{r}^2 + v_{\phi}^2)]}^{3/2}}\,. 
\label{angmom}
\ENA

\noindent
Here $A = 4\pi G^2M_*^2\rho(r)$ represents the coefficient of drag force $\vec{F}_{d,n}$ in the direction 
of $\hat{n}$, where $\vec{F}_{d,n} = (A/v^2)\hat{n}$. The enclosed mass $M_{\rm enc}(r)$ is integrated 
over the density regime as
\EQ
M_{\rm enc}(r) = 4\pi r^2 dr \rho(r) = 4\pi \rho_0 r_{0}^3 \int
\frac{(r/r_0)^2 d(r/r_0)}{(1 + r/r_0)^{2.2}} dr.
\label{mencprof}
\EN
 
\noindent 
As mentioned before, the dynamics of the protostars following the above equations is calculated until 
$r \approx 2\pc$, and the final mass is evaluated at this point. Protostars, once escaping this region, 
will no longer be part of the primordial cluster.  

\subsection{Numerical scheme}

We start our semianalytical calculations from the set of initial conditions described in Section~\ref{subsec:initial}, 
along with the equations of motion and angular momentum conservation of protostars.
To this effect, we used the standard Runge--Kutta fourth-order method to solve the system of Equations (5) - (8) 
with compressible flow coupled with gravity and drag in spherical-polar coordinates. However, solving the set of 
differential equations closest to the central core requires ``a much smaller'' time step because of the large
density gradient of the ambient gas. Therefore, the time step should be chosen in a way that allows us to 
probe the details of the dynamics near the center, as well as far away, where the density gradient is less. The 
fixed time step, even if it is very small, can produce incorrect results. In our simulation, we have therefore 
used an adaptive time-stepping scheme, calculated from the velocity and accretion of protostars, which 
satisfies the well-known Courant--Friedrichs--Lewy (CFL) condition. In other words, the system should evolve
both in the radial and the azimuthal direction with the time step $\Delta t_r$ and $\Delta t_{\phi}$ respectively, 
and will converge with `minimum time step' ($\Delta t_{\rm min}$), 
where $\Delta t_{\rm min} \equiv \,{\rm min}(\Delta t_r \,, \Delta t_{\phi})$. The time step ($\Delta t$) is given by 
\EQ
\Delta t_r = \varepsilon \Bigl|\frac{v_{r, i}^2}{dv_{r, i}/dt}\Bigr| \,\,\,\,{\rm and} \,\,\,\, \Delta t_{\phi} 
= \varepsilon \Bigl|\frac{v_{\phi, i}^2}{dv_{\phi, i}/dt}\Bigr| \nonumber
\EN

\EQ
\Delta t = {\rm min}\Bigl[\varepsilon \Bigl(\frac{v_{r, i}^2 + v_{\phi, i}^2}{(dv_{r, i}/dt)^2 + (dv_{\phi, i}/dt)^2}\Bigr)^{0.5} \Bigr] 
\EN

\noindent 
Here $\varepsilon \le 5\times 10^{-3}$ is the dimensionless Courant number that converges the system 
satisfactorily. This gives high accuracy to the dynamics of both types of protostars: those that escape through 
ambient gas, and others that orbit around the gravitationally bound system and continue to accrete mass. 
The time step in this study is also comparable to those used in previous simulations with the 
{\sc Gadget}-2 SPH numerical code \citep{dnck15}. 

Note that in a full 3D simulation it is extremely difficult to probe the entire dynamics with higher accuracy in 
smaller time steps. It is also computationally expensive to follow the trajectory and evolution of each protostar 
and its interaction with the ambient gas. Thus, apart from allowing for a wide parameter space to be explored, 
our semianalytical approach offers an alternative route compared to the full-scale 3D simulations. The initial
time corresponds to the time of formation of protostars. For those protostars that are able to escape the cluster, 
we follow their evolution until the time at which their distance from the center is $\sim 2\pc$. For other 
protostars that are unable to escape the cluster, we continue to follow their evolution until they accrete 
significant mass. 

\subsection{Correctness of the numerical model}

At this point it is important to examine whether the results from the semianalytical model are accurate. 
We have therefore solved the same set of equations (see Equations (5) - (8)) in a Cartesian coordinate system 
so as to verify the validity and correctness of the performance of the numerical code in the spherical-polar 
coordinate system from which the results presented in the next section have been generated. Moreover, 
we have cross-checked the results obtained from both codes with analytical calculations, which one can 
verify after employing certain physical assumptions. Depending on the size of the core ($r_0$), one can 
find a distribution function for the estimated escape velocity (Equation~(\ref{escape})) that ranges from 
approximately $9$ to $15\kms$, in agreement with results from cosmological simulations. As an example, 
in Section~\ref{subsec:initial} we have calculated the escape velocity $\sim 11 - 13\kms$, which is 
very close to the actual value, and which provides the escape speed exactly as has been seen in 
cosmological simulations \citep{greif12,johnson15}. 

%%%%%%%%%%%%%%%%%%%%%%%%%%%%%%%%%%%%%%%%%%%%%%%%%%%%%%%
\section{Results}
\label{sec:result}

%%%%%%%%%%%%%%%%%%%%%%%%%%%%%%%%%%%%%%%%%%%%%%%%%%%%%%%
\begin{figure*}
\centerline{
\includegraphics[width=3.6in]{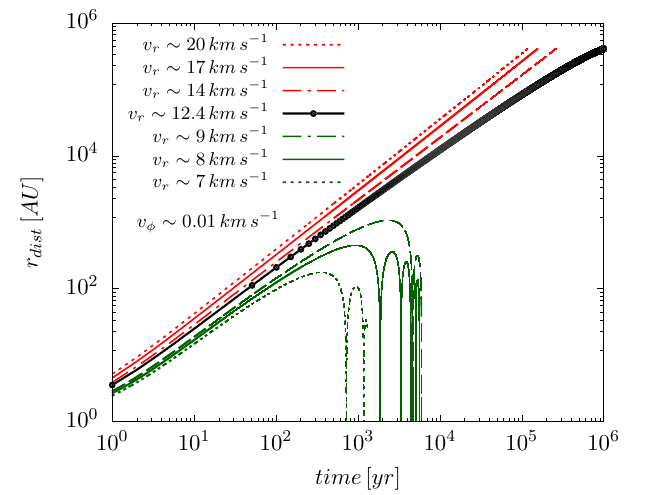}
\includegraphics[width=3.6in]{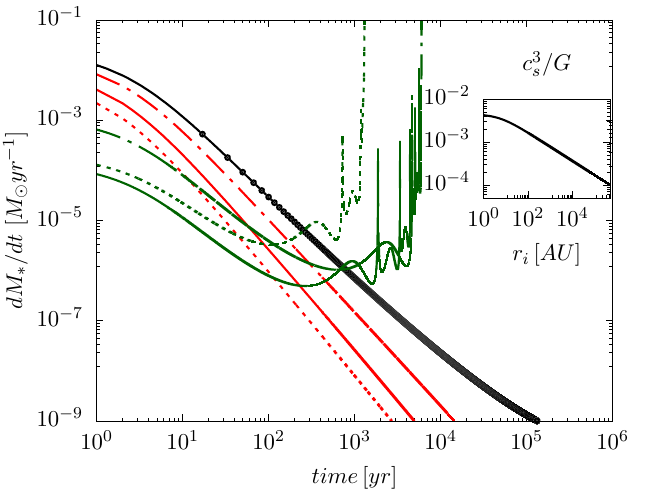}
}
\caption{\label{trajectories} Time evolution of the protostar trajectories (left panel) and mass accretion 
rate (right panel), plotted for two cases: one in which the initial radial velocity is larger (red lines) than the 
escape velocity $v_{esc} \approx 11 - 13\kms$ and another in which it is smaller (green lines). 
The size of the cluster is around $\sim 2\pc$. The core density and core radius are $n_{0} = 10^{13}\cmn$ 
and $r_{0} = 5\au$, respectively. Protostars with $v_r > v_{esc}$ and certain critical masses are ejected from 
the cluster at around a million years with a final mass $\le 0.8\msol$. In contrast, protostars with speed less than 
the escape speed move around the ambient gas against gravitational drag and accrete a sufficient amount of gas 
to end up becoming massive stars. As expected, the mass accretion rate could be much larger near the 
central regime (low angular momentum) and thereby approaches $\sim c_s^3/G$ (represented by the radial 
variation shown in the inset of the right panel), whereas it drops significantly for the ejected protostars at the outer 
regime, where density is comparatively lower following the $\rho\,\propto\,r^{-2.2}$ profile.\\}
\end{figure*}
%%%%%%%%%%%%%%%%%%%%%%%%%%%%%%%%%%%%%%%%%%%%%%%%%%%%%%%%

In this section, we examine the dynamical time evolution of the trajectories and mass accretion by the protostars 
that move around the ambient medium inside the gas clumps. 

\subsection{Dynamical evolution}
\label{fixnoro}

The evolution of protostars with speed on either side of the escape speed is shown in Figure \ref{trajectories}, 
where the initial core density and core radius are $n_{0} \sim 10^{13}\cmn$ and $r_{0} \sim 5\au$, 
respectively. The trajectories of the protostars having initial radial velocities ($v_{r}$) for a particular choice of 
the azimuthal velocity $(v_{\phi})$ are plotted as a function of dynamical time (left panel of Fig.~\ref{trajectories}). 
As discussed in Section~\ref{subsec:initial}, the escape speed has a distribution function that depends on the 
position of these protostars in the cluster. For primordial clumps, we estimate the protostars to escape at a
speed of nearly $v_{\rm esc} \approx 11 - 13\kms$. In the figure, the black solid line represents the protostar 
that has a radial velocity roughly equal to the escape speed $\sim 12.35\kms$ corresponding to the azimuthal 
velocity $v_{\phi} \sim 0.01\kms$. The red lines correspond to protostars with speed $\sim$20 (dotted), $17$ 
(solid), $14$ (dash--dotted) and green lines correspond  to $9$ (dash--dotted), $8$ (solid), and $7$ (dotted) 
in units of $\kms$, respectively. We keep the azimuthal velocity comparatively small because 
some of the fragments may have lower angular momentum \citep{stacy13}. We notice that some of the 
protostars (red lines) can directly travel the path of $\sim 2\pc$ at around a million years. For these protostars, 
the initial masses ($m_{i}$) were chosen in such a way that the final mass $m_{f} \leq 0.8\msol$ at the time 
of escaping the cluster. Furthermore, we also find that different initial radial velocities lead to subtle deviations. 
For instance, protostars with a larger radial component result in a slightly steeper curve and therefore leave 
the cluster earlier. In contrast, protostars with speed $v_r \ll v_{esc}$ (green lines) keep on orbiting around the 
central clumps while being slowed down by the gravitational drag from the ambient gas density.  For these 
stars, the initial masses were chosen in the range $m_{i} \sim 0.01 - 0.3\msol$. 

In this context, it is important to analyze the mass accretion rate and effect of drag force (parameterized 
by '$A$' shown in Equations~(\ref{motion})--(\ref{angmom})) on the dynamical evolution of the protostars 
and ambient gas that are trapped inside the gravitational well of the cluster. The gravitational drag can 
change the trajectories of the orbiting protostars, as is the case for X-ray binaries \cite[see, e.g.,][for further 
discussion on the orbital motion of protostars inside a protocluster]{edgar04}.
The right panel of Figure~\ref{trajectories} shows the time evolution of the mass accretion rate of all protostars 
computed from our semianalytical model. Mass accretion near the core is controlled by the self-gravity of the 
collapsing gas and hence is expected to be proportional to the theoretically predicted value $\sim c_s^3/G$ 
(the radial variation of which for $n_{0} \sim 10^{13}\cmn$ and $r_{0} \sim 5\au$ is shown in the inset).  

Once we start our semianalytical run with the adaptive time-stepping scheme, protostars immediately 
accrete a substantial amount of ambient gas ($dM_*/dt\,\sim\,10^{-4}-10^{-2}\,{\rm M_{\odot}\,yr^{-1}}$) at 
the very initial epoch of time. In this regime the Bondi--Hoyle accretion rate is roughly of the same order as 
the estimated $c_s^3/G$. However, as the protostars of interest in the present study are those that move 
to the outer envelope, where density is low (i.e., $n_0\,\sim\,10^3-10^4\cmn$), the Bondi--Hoyle accretion 
onto the ejected protostars (red and black lines) gradually becomes very low (of the order of 
$10^{-9}\msol\yr{-1}$). This also reflects the fact that 
mass evolution of the ejected protostars remains nearly constant over time, so that the final mass at the time 
of escaping the cluster remains as low as $0.8\msol$. On the other hand, protostars that keep orbiting the 
core region (low angular momentum) tend to have larger accretion rates compared to the ones that escape 
the cluster, after about a thousand years into their evolution. 

%%%%%%%%%%%%%%%%%%%%%%%%%%%%%%%%%%%%%%%%%%%%%%%%%%%%%%%%
\begin{figure}
\centerline{
\includegraphics[width=3.6in]{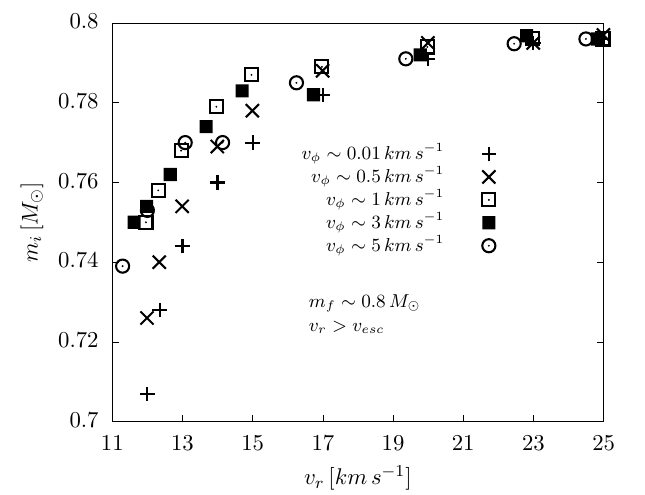}
}
\centerline{
\includegraphics[width=3.6in]{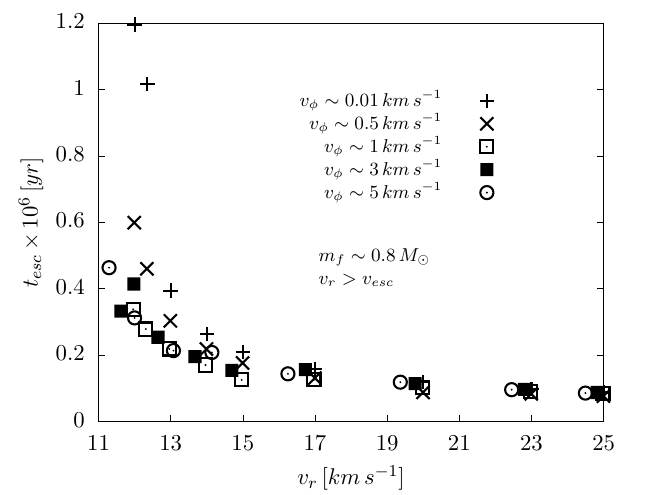}
}
\caption{\label{mass_esc} Stellar initial mass (top panel) and the escape time (bottom panel) of the 
protostars before leaving the cluster are plotted as a function of radial velocity, $v_r$, for different values 
of initial azimuthal velocity, $v_\phi$. This shows that the protostars with a mass range 
$m_i \sim 0.7\hbox{--}0.79$ can be ejected with $v_r > v_{esc}$ while experiencing the gravitational drag 
from the surrounding gas, and in the process they accrete a negligible amount of mass to end up becoming 
low-mass stars with $m_f \approx 0.8\msol$ (as shown in Figure~\ref{trajectories}).\\  
}
\end{figure}
%%%%%%%%%%%%%%%%%%%%%%%%%%%%%%%%%%%%%%%%%%%%%%%%%%%%%%%%%

Next, we explore the effect of rotation on the mass accretion phenomenon for the ejected protostars. 
In Figure~\ref{mass_esc} we show the distribution of the initial masses of the protostars ($m_{i}$) and the 
computed escape time ($t_{\rm esc}$) taken by the evolving protostars to reach a distance $r \ge 2\pc$ 
in the top and bottom panels, respectively. We consider protostars having different initial values of 
$v_{\phi} \sim 0.01, 0.1, 0.5, 1, 3$ and $5\kms$, as a function of initial radial velocities in the range 
$v_{r} \sim 11 - 25\kms$. The top panel shows that up to $v_r \sim 17\kms$, the required initial mass 
increases for higher $v_{\phi}$, given a final mass that is close to $m_f \sim0.8\msol$. 
An initial high speed $v_{r} \gg v_{esc}$ (denoted by red lines in Figure \ref{trajectories}) ensures that the 
mass accretion is relatively small before the protostar escapes the cluster. Accordingly, we chose the initial 
masses of these protostars to be larger than $\sim 0.78\msol$, if they are to end up with a final mass less than 
$0.8\msol$. For initial velocity just over the escape velocity, we find that the initial mass may need to be as 
low as $0.65\msol$ (not shown in the figure). Thus, the net accretion can be up to about $20\%$ for protostars 
that escape. The bottom panel shows that for $v_r \leq v_{esc}$, the time taken to escape from the cluster 
decreases as $v_{\phi}$ increases from $\sim 0.01$ to $\sim 1\kms$. On the other hand, for protostars that 
start with $v_r \gg v_{esc}$, we find that the escape time is independent of the initial azimuthal velocity of the 
protostar. Although the focus of our work is on the dynamical evolution of those protostars that are able to 
escape from the cluster, we comment here on mass evolution of the other protostars that have $v_{r} \ll v_{esc}$. 
As evident from the green curves in Figure~\ref{trajectories}, these protostars continue to remain inside 
the cluster. Since we do not include any explicit feedback scheme in our semianalytical model, we find 
that these protostars continue to accrete and gain in mass uninhibitedly, making it impossible to predict 
their final masses. However, until the time we are able to follow their evolution, we find that these protostars 
acquire a substantial amount of mass in the range of $\sim 1 - 20\msol$ or even more depending 
on the initial conditions of the cluster.

Further, we find that the mass accretion coincides with initial deceleration as can be predicted from the 
trajectories, i.e., the number of times that these protostars rotate, as shown in Figure \ref{trajectories}. The 
accretion rate is expected to be higher if the protostar is moving with a slower velocity and/or through 
denser medium. Therefore, for $v_{r} < v_{esc}$, the protostar accretes a large amount of mass relative 
to its initial mass and can even increase its mass by an order of magnitude by the time the orbit is 
significantly affected by the gravitational drag. They can therefore continuously accrete mass to end up 
becoming massive stars, or may merge with another star 
\cite[see recent state-of-the-art simulations by][for the evolution of supermassive Population III stars]{haemmerlE18}. 
Although the final mass distribution of these massive protostars needs a more careful investigation 
and is beyond the regime of interest for this study, we would like to mention that our result for these 
protostars is consistent with the prediction that the first stars have masses with 
$\sim 10 - 1000\msol$ \citep{hirano14}. 
Thus, even if the initial mass function of protostars does not contain very massive stars, accretion can 
lead to a substantial increase in mass. Such stars can potentially lead to supernovae \citep{whalen14} 
and also result in early formation of black holes \citep{g15,umeda16}. 
%%%%%%%%%%%%%%%%%%%%%%%%%%%%%%%%%%%%%%%%%%%%%%%%%%%%%
\begin{figure}[h]
\centerline{
\includegraphics[width=3.6in]{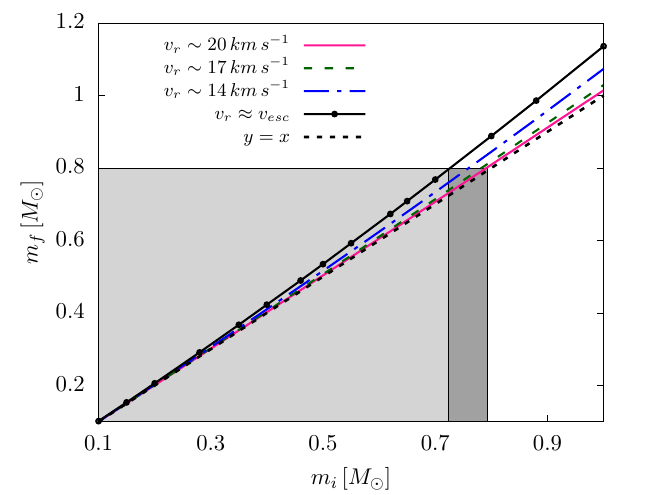}
}
\caption{\label{mimf} 
Growth in mass as a function of the initial mass are plotted for protostars that escape the 
gravitationally bound system for initial masses in the range $m_{i} \sim 0.1 - 1\msol$. 
The mass accretion is higher for a lower radial velocity. The dark-gray regime denotes the 
protostars having survival possibility while leaving the cluster with a final mass 
$m_f \approx 0.8\msol$, whereas the light-gray regime denotes the protostars with a mass 
$m_f \ll 0.8\msol$.\\}
\end{figure}
%%%%%%%%%%%%%%%%%%%%%%%%%%%%%%%%%%%%%%%%%%%%%%%%%%%%%%

In figure~\ref{mimf} we plot the growth in mass as a function of the initial mass for all protostars that escape 
the gravitationally bound cluster. We consider initial masses in the range $m_{i} \sim 0.1 - 1\msol$. 
As mentioned earlier, the final mass ($m_f$) is the mass of the protostar measured once the protostar 
reaches $r \approx 2\pc$. Each curve corresponds to a different initial radial velocity. We find that the mass 
accreted is higher for a lower initial $v_{r}$. This is to be expected, as the accretion rate is higher for a lower 
velocity and in such a case the protostar spends more time in the core region, where the ambient density 
is higher. Further, we note that for a given radial velocity the mass accreted is higher for a higher initial 
mass. This again relates directly to the increased accretion rate. As the radial velocity increases, the final 
mass vs initial mass curves come closer to the $y=x$ line shown for reference. Note that the increase in 
mass for protostars that escape the system is small and is limited to $\sim 0.2\msol$. Therefore, this process, 
i.e., increase in accretion rate near the center of the core, cannot be responsible for a significant increase in 
the mass of protostars that escape the cluster where they are born.

Of course, our semianalytical approach still leaves out processes that are important in a physically realistic 
scenario. Radiative feedback and thermal feedback are likely to evaporate the core of the clumps, and hence 
the increase in mass will be expected to be comparatively less. Nevertheless, it is heartening to note that 
the final masses obtained from our simple model are comparable to that predicted in other simulations 
\cite[e.g.][]{greif12,stacy13,hirano14}. The increase in mass is higher for a lower initial radial velocity, as 
expected from the expression for the accretion rate. This confirms the fact that the Bondi--Hoyle accretion 
flow is the key process in early evolution of protostars inside the cluster.

%%%%%%%%%%%%%%%%%%%%%%%%%%%%%%%%%%%%%%%%%%%%%%%%%%%%%%
\subsection{Dynamical evolution for different choices of $n_{0}$ and $r_{0}$}
\label{varnoro}

%%%%%%%%%%%%%%%%%%%%%%%%%%%%%%%%%%%%%%%%%%%%%%%%%%%%%%%
\begin{figure*}
\centerline{
\includegraphics[width=3.6in]{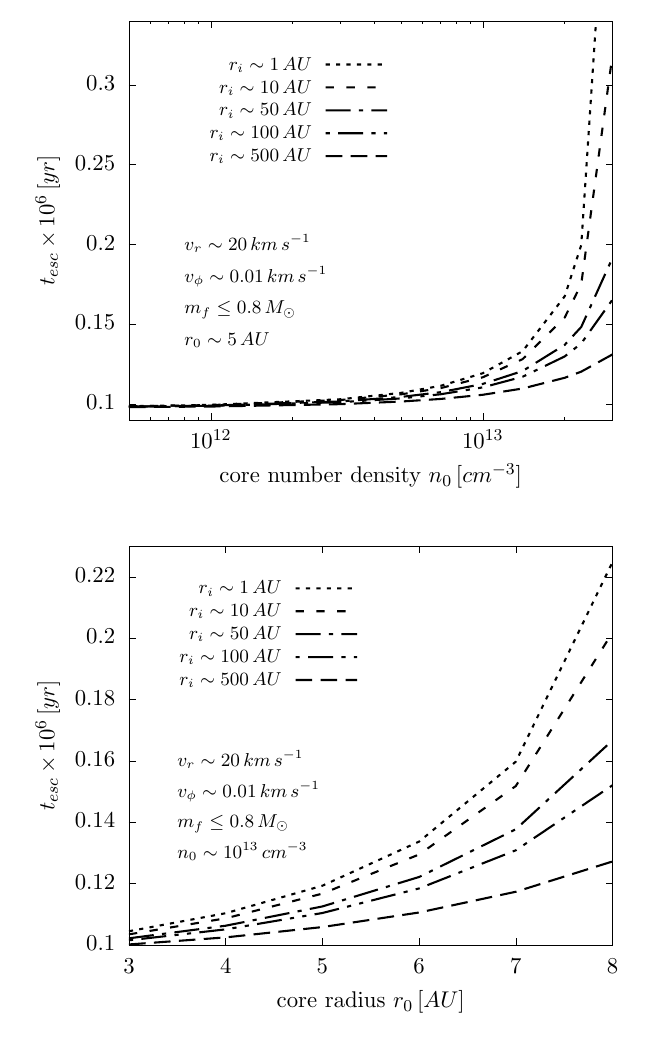}
\includegraphics[width=3.6in]{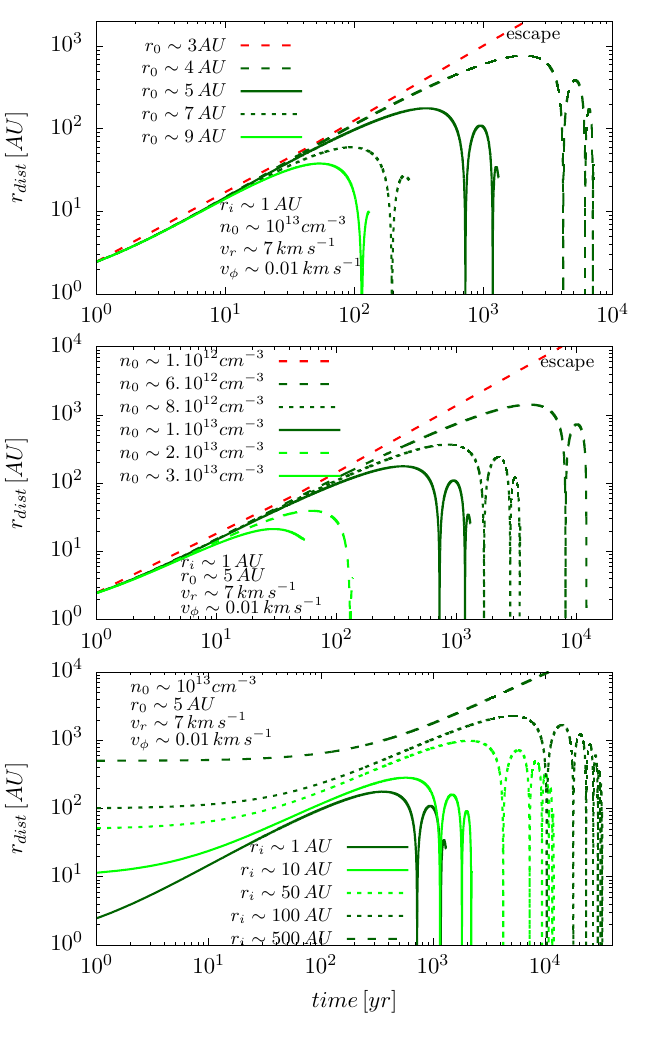}
}
\caption{\label{core} Left column: escape time ($t_{\rm esc}$) of the ejected protostar having 
initial $v_r \sim\,20\kms$, are shown as a function of both central ambient density 
(represented by the number density $n_0\,[{\rm cm^{-3}}]$ in the top panel) and size of the 
central regime (represented by the core radius $r_0\,[{\rm au}]$ in the bottom panel). 
Each line represents the initial position of the protostar that varies over a wide range 
around $1--500\au$, while other parameters such as rotational velocity and initial mass 
of the protostar remain the same. 
Right column: the three panels show how the trajectory of a protostar with initial 
$v_r \sim\,7\kms$ is affected during the evolution of core radius (top panel), evolution 
of central density (middle panel), and for a different initial distance of the protostar 
(bottom panel). There is a possibility that the protostar can even escape the cluster 
provided that the core is small enough with $r_0 \le\,3\au$ and $n_0 \le\,10^{13}\cmn$.\\}
\end{figure*}
%%%%%%%%%%%%%%%%%%%%%%%%%%%%%%%%%%%%%%%%%%%%%%%%%%%%%%%%

Until now, we discussed the dynamics of the protostars that are initialized with a wide 
range of velocities, masses, and positions when the central density attains a maximum value 
of $n_0\sim\,10^{13}\cmn$ within a size of $r_0 \sim\,5\au$. However, in reality the ambient 
density is expected to be changing with time mostly in the central core, and hence the size 
of the core also evolves with time. 
In this context, early semianalytical works by \citet{shu77} and \citet{suto88} studied the 
evolution of the central protostar using 1D models of cloud collapse. The results 
obtained from these works suggest a time-dependent evolution of $\rho(r,t)\propto t^{-1/2}r^{-3/2}$ 
for $\gamma = 1.09$ in the very central regime\footnote{In the notation used in their papers, 
this correspond to solutions for $x \ll 1$ where $x$ is the dimensionless similarity variable.}. 
In these calculations, the central protostar expands as it accretes matter from the surrounding 
low-density material, with the head of the expansion wave traveling outward at the speed of 
sound ($r = c_{\rm s}\,t$). Beyond this radius, the density profile switches to the familiar 
time-independent $\rho(r) \propto r^{-2.2}$. Although the dynamics discussed in these works is 
well understood, it sharply differs from results of recent 3D cosmological simulations
that follow the evolution beyond the formation of the central protostar. In these 
simulations, the behavior of the decreasing core density as implied from the analytical 
$\sim t^{-1/2}$ scaling is not seen \citep[see][]{greif12,bgsh15}. In contrast, they clearly 
show that the core density maintains a constant profile for $r < 1$ au and also increases 
from $\sim 10^{18}$ to a few times $\sim10^{21}\cmn$, maintaining the well-known 
$r^{-2.2}$ profile in the outer region. In an idealized simulation that focuses on the 
evolution of a single protostar, \citet{md13} show the existence of an expanding shock 
front (at the surface of the central protostar) inside which the core density maintains a 
$\rho \propto r^{0}$ profile\footnote{This cannot be captured in the 1D models of 
\citet{shu77} and \citet{suto88}.}. The decrease in density implied by the works of 
\citet{shu77} and \citet{suto88} occurs in a region immediately outside the expansion 
radius, while the density remains $\propto r^{-2.2}$ farther out. 

Motivated by the above results, we explore in this subsection how the dynamics of the 
protostars evolve for a range of values of $n_{0}$ and $r_{0}$ of the evolving ambient 
gas with particular focus on protostars that start with $v_{r} \sim 20\kms > v_{esc}$ and 
thus are likely to escape from the cluster. However, for the sake of completeness, we also 
explore the dynamical evolution of protostars that start with $v_{r} \sim 7\kms < v_{esc}$. 
The results are shown in Figure~\ref{core}.

In the left column, we plot the computed escape time ($t_{\rm esc}$) of the protostars as they 
travel through the varied dense regime, i.e., for different core density (top panel) and for 
different core radius (bottom panel). The curves with different line styles in each panel denote 
protostars that start out at varying initial distances ($r_{i}$) from the central core. Other 
characteristic parameters such as the initial mass ($m_{i}$), $v_r \sim 20\kms$ and 
$v_{\phi} \sim\,0.01\kms$ are kept fixed such that the final mass $m_{f} \le 0.8\msol$. 
It is clear from the figure that for $n_{0} \leq 2\times 10^{12}\cmn$ the escape times of the 
protostars having different initial $r_{i}$ are the same, $\sim 0.1\,{\rm Myr}$. Beyond this and 
with increasing $n_{0}$, protostars that start out closer to the core experience a substantial 
amount of drag owing to stronger gravity. This results in them taking more time to escape from 
the cluster. This effect is more pronounced for densities, $n_0 \ge\,10^{13}\cmn$, which also 
represents the critical density for 'sink' formation obtained in recent numerical simulations 
\cite[e.g.,][]{cgsgkb11,stacy13,dnck15,dutta16a}. On the other hand, keeping $n_{0}$ and 
other parameters fixed, the different curves in the bottom panel show that the escape time 
increases as the core radius is increased from $3$ to $8$ au. Further, for $r_{0} > 3\au$, 
protostars that are placed far away from the core (e.g., $r_{i} \sim 500\au$) escape earlier 
than those that are placed closer to the centre. 

In the right column, we show the evolution of protostars for an initial $v_{r} < v_{esc}$ and 
$m_{i} \approx 0.03\msol$. Here we first consider protostars placed at $r_{i} \sim 1\au$ with 
$n_{0} \sim10^{13}\cmn$, $v_{r} \sim 7\kms$ and $v_{\phi} \sim 0.01\kms$, for different choices 
of the core radius (top panel). We find that the only instance where a protostar can escape 
corresponds to a core radius $r_{0} \sim 3\au$. In all other cases, the protostars are unable to 
escape the cluster, with the effect of gravitational drag becoming more severe with increasing 
values of $r_{0}$. In the middle panel, the trajectories are plotted for six different values of 
$n_{0}$ for $r_{0} \sim 5\au$ and $r_{i} \sim 1\au$. Once again, we find that for very low values 
of $n_{0} \sim 10^{12}\cmn$ a protostar placed very close to the centre can escape the cluster. 
However, for increasing core densities protostars are unable to escape the cluster. In particular, 
protostars that evolve in a more denser regime experience a strong gravitational drag and hence 
are able to move a distance of only few tens of au before being stopped by the gaseous medium 
(light-green dashed and solid curves). This is similar to the effects seen for higher core size 
(i.e., $r_0\,\ge\,7\au$ in the top panel). The bottom panel shows the trajectories for protostars 
placed at a different initial distance from the central region. We again find that the frequency at 
which a protostar rotates can be affected by its initial position. To summarize, there exists 
a possibility for a protostar to escape the cluster even for $v_r\,\sim\,7\kms$ provided that the 
central core density and core size are comparatively low enough (red line). This can easily 
be explained from the escape speed (Equation~\ref{escape}) that depends on both $r_0$ and 
$\rho_{0}$. However, this is rather a special case obtained from our simple semianalytical 
model. In realistic situations such a scenario is unlikely, as the number density is always 
expected to be of the order of $\sim\,10^{13}\cmn$.  

It is to be noted here that we have assumed the core density to remain constant with 
a radial profile given by Equation~(\ref{denprof}) during the evolution of the protostars. 
Similar to other 1D collapse models \citep{shu77, suto88, maki07}, our semi-analytical 
model cannot capture the existence of an expanding shock front, which would require
a full 3D numerical simulation that is beyond the objectives of the present 
study. This would be particularly necessary if one were to study the trajectories of 
protostars that are unable to escape the cluster. On the other hand, protostars that start 
with $v_{r} \gg v_{esc}$ and are initially located farther out from the central core are unlikely 
to be affected by the change in radial profile of the central protostar.
Thus, with the help of the semianalytical presented here, we can only begin to have a 
preliminary idea of how the evolution of both the core size and core density can influence 
the trajectory and fate of the ejected protostars. 

%%%%%%%%%%%%%%%%%%%%%%%%%%%%%%%%%%%%%%%%%%%%%%%%%%%%

\section{Summary and Discussion}
\label{sec:summary}

In continuation of an earlier study \citep{dutta16a}, our aim here was to investigate the evolution of 
the protostars that formed out of fragmentation of unstable circumstellar disks of rotating primordial gas 
clumps. To this end, we developed a semianalytical model using the well-known Bondi--Hoyle accretion to 
follow the dynamics of the evolving protostars and their interaction with the ambient medium. The initial 
conditions of our model, such as the density and temperature profile of the ambient gas, mass, and 
radial and rotational velocities of the newly formed protostars, are similar to the physical properties of 
the ambient gas and protostars reported in more complex, cosmological simulations of primordial star 
formation. We solved the governing equations (see Equations (5)--(8)) in a spherical-polar coordinate system 
using Runge-Kutta fourth-order method with an adaptive time-stepping scheme that satisfies the CFL 
condition. We have also included the standard drag force experienced by these protostars while moving 
around the ambient medium. The implementation of this approach thus provides a better understanding 
of the dynamical system evolving for a longer period of time. 

Using the above methodology, we have explored two possible scenarios that are likely to happen 
depending on the initial configuration of the gas clumps. For example, there is a high probability that 
some of the primordial protostars can merge below a certain fragmentation scale \citep{hosokawa16,hb17} 
or even collapse further to form compact objects \citep{g15,latif18}. On the contrary, a few recent studies 
have demonstrated that a number of low-mass Population III stars are ejected from the gravitationally bound 
cluster with radial velocities larger than the escape speed of the cluster \citep{greif12,johnson15,ishiyama16}. 
Indeed, high-resolution {\sc Gadget-2} SPH simulations of \citet{dutta16a} also found that a fraction of 
these protostars can overcome the gravitational drag force and simultaneously accrete only a small 
amount of mass before being ejected. However, because of computational limitations, it was not possible 
in 3D simulations to track their dynamics and interaction with the surrounding gas. As a number of such 
protostars have mass below $0.8\msol$, it is likely that they are still on the main sequence at the present 
time. Ejection can also happen if the primary, more massive star goes supernova and the gas clump loses 
much of its mass \citep{komiya16}, though we have not considered this process here. Based on the results 
obtained from our semianalytical model, we summarize below the main findings of our work. 

\subsection{Key points}

We first focused on the dynamics of the evolving protostars where the central density attained a 
maximum value of $n_{0} \sim 10^{13}\cmn$ and core radius $r_{0} \sim 5\au$. For a given initial 
{\bf $v_{\phi} \sim 0.01\kms$}, time evolution of the protostar trajectories as shown in Figure.~\ref{trajectories} 
reveals that if the initial $v_r \gg v_{esc}$, the protostars can escape the gravitationally bound system in 
$\sim 10^{6}\yr$, with final masses $\leq 0.8\msol$. Evolution of protostars for different initial values 
of $v_{\phi}$ shows (see Figure~\ref{mass_esc}) that up to $v_{r} \sim 17\kms$ the required initial mass 
increases for higher $v_{\phi}$, given a final mass $m_{f} \sim 0.8\msol$. The same figure shows 
that for $v_{\rm r} \leq v_{\rm esc}$ the escape time decreases with the increase in $v_{\phi}$ from 
$\sim 0.01$ to $\sim 1\kms$, compared to the case when $v_{\phi} \sim 3, 5\kms$. However, for 
$v_{r} > 17\kms$, the escape time becomes independent of the initial azimuthal velocity. Figure~\ref{mimf} 
shows that mass accretion by the protostars increases as we go to lower initial radial velocities for 
those that escape the cluster. For a given initial speed and radial motion, the final mass is higher for 
those with a higher initial mass. The lowest probable mass range of the protostars from our model is 
comparable to the recent results from exciting observations by \citet{schlaufman18}. Protostars with 
$v_{r} < v_{esc}$ are unable to escape the cluster. They remain in the cluster, spend considerable 
time in the core region, and continue to accrete from ambient gas. In the absence of realistic feedback 
processes in our semianalytical model, this leads to a large enhancement in mass over time. 
During the time in which we have followed their evolution, we have found that these protostars can 
accrete enough mass to grow up to $10 - 20\msol$, or even more. We also expect the protostars to pick 
up high velocity in encounters \citep{aarseth98}.

Next, we probed the dynamics of the protostars for different initial core densities and core radii. Our 
intention here was to explore the evolution of trajectories of protostars placed at different initial distances 
from the central core. To this end, we focused on two extreme choices of the initial $v_{r}$, one larger 
and the other smaller than the escape speed. We found that for $v_{r} \sim 20\kms$, the time 
required to escape the cluster increases with the increase in core density. In particular and as expected, 
protostars that are placed closer to the center require much more time to escape the cluster compared to 
those that start their evolutionary journey far away from the central core. Similar inferences can be drawn 
for the variation of the escape time with core radius. Here we kept the core density fixed at 
$n_{0} \sim 10^{13}\cmn$ and found that the escape time increases as the core radius drops from 
$3$ to $8\au$. In the other case where initial $v_{r} \sim 7\kms$, we found that it is possible for a 
protostar placed very close to the central core to escape, provided that either the core density is low 
($n_{0} \sim 10^{12}\cmn$), or the core radius is small ($r_{0} \sim 3\au$). For all other choices of 
$n_{0}$ and $r_{0}$, protostars are unable to escape the cluster. At very large core densities or for 
large core radius, protostar trajectories experience strong gravitational drag, resulting in them being 
able to move a distance of only a few tens of au.

\subsection{Caveats}

The main objective of this paper was to probe whether the mass
  accretion phase and the associated dynamical evolution of protostars
  can be described by a simple semianalytical model of Bondi--Hoyle
  accretion.
  Here we enumerate the caveats of our work to clarify the context
  within which the semianalytical model and our results may be
  considered reliable.

  \begin{itemize}
  \item
    We have assumed that the core density ($n_{0}$) and the core
    radius ($r_{0}$) remain unchanged during the evolution of the
    protostars.
    In reality, both these parameters are likely to change owing to
    ongoing accretion onto the protostars and due to changes in
    temperature and density distribution of the circumstellar disk
    system \cite[as shown in][]{on98,yoh08,cgsgkb11,dutta16a}.  
    Computational limitations imply that this has not been
    studied at the required level in simulations, in particular the
    evolution of density in the central region over a period of
    $10^6$ yr has not been explored.
    However, as we have shown, for protostars with $v_{r} > v_{esc}$,
    this does not lead to significant uncertainty in the results.
    This also implies that our semianalytical model cannot be used
    in its present form for studying the evolution of protostars that
    spend significant time in the central regions.
  \item
    We have not included any radiative feedback in our model.
    A number of previous work have shown that feedback effects are a
    crucial factor in determining the final mass of protostars.
    These effects become important over time scales of $\sim 10^{4}\yr$
    after the formation of the first protostar \citep{hosokawa16,sbl16}.
    Radiative feedback raises the temperature and lowers the density of 
    gas in the central region, which lowers the accretion rate at late times
    for stars in this region.
    In the present study our focus is on stars that escape the gas
    clump where these are formed.
    A close look at Figure~\ref{trajectories} shows that protostars with
    initial $v_{r} > v_{esc}$ are able to travel to distances of $\sim
    2\pc$ within a few times $10^{5}\yr$.
    Thus, such protostars spend very little time in the central
    regions, and the absence of radiative feedback does not affect our
    results significantly.
    Radiative feedback needs to be taken into account if we study
    stars in the central region.
    However, probing the fate of such stars remains outside the scope
    of this study.
  \end{itemize}

In a nutshell, notwithstanding the limitations of the semianalytical model, our results imply that 
it is plausible that the first stars might have formed in a broad range of masses during disk 
fragmentation -- fragments that escape the cluster have the possibility to enter the main sequence 
as low-mass Population III protostars ($\le 0.8\,{\rm M_{\odot}}$) and hence could survive until the 
present epoch of time. On the other hand, protostars that remain inside the cluster can end up evolving 
as massive stars ($\sim$$1 - 20\,{\rm M_{\odot}}$) depending on the initial conditions and their 
accretion history. 

\subsection{Observation of Metal-poor stars}

Understanding the implications of our work for observations requires us to make assumptions about 
the initial mass function of Population III stars. In principle, this can be constrained using observations 
of metal poor halo stars, and in this section we comment on this issue. 

Observations of metal poor stars in the halo of the Galaxy suggest a high floor of around $0.5\msol$, 
i.e., all known low-metallicity stars have masses higher than this threshold. The process of accretion 
in the parent cloud studied here can lead to an increase in the mass by a small amount. However, 
stars that escape the parent cloud experience only a very small mass increase. This on its own 
cannot explain the absence of very low mass stars. The increase in mass for stars that remain in the 
cloud is fairly large, and hence these stars are not expected to survive to the present epoch. 
In future studies, we plan to consider evolution of a group of stars, and it is possible that many-body 
interactions lead to a more nuanced understanding. 

The issue of whether some of the Population III stars may have survived until today and the possible sites of 
where such stars are likely to be found has been explored by a number of authors in the past 
\citep{bond81,salvadori10,ishiyama16,tanaka17}. Some recent studies suggest that the oldest Population III 
remnants should be spread throughout the entire Galaxy \citep{scannapieco06,brook07}. Others predict 
that Population III survivors are likely to be concentrated toward the Galactic bulge 
\citep{dms05,tumlinson10,bsw15}. Finally, \citet{magg18} extended these studies to find that low-mass 
Milky Way satellites are more likely to contain Pop III stars than the Milky Way itself, and that low-mass 
satellites will serve as promising targets in the search for Population III survivors. However, to estimate the rate 
of Population III detection for future surveys, it may be more worthwhile to concentrate on determining the 
number and initial mass function of these stars \citep{desouza14,deb17,griffen18,stf18}.

Although previous work suggests that Population III survivors can reside in both the bulge and halo, a more 
careful theoretical approach is needed to settle the issue. In this regard, our work shows that the lowest-mass
protostars are the ones that escape the cluster of formation, indicating that these are poorly bound 
to their host halos. It has been shown that in mergers of halos the most tightly bound objects end up 
near the centre of the merged halo, whereas loosely bound objects in the parent halos end up on the 
outskirts \citep{sw98}. In view of the preserved ordering by binding energy in mergers, we expect that 
these stars are more likely to be found in the halo or outer parts of the bulge. On the other hand, stars 
that do not escape go on to accrete enough mass to eventually explode as a supernova. Such stars can 
potentially be seeds for black holes.

\subsection{Upcoming work with data analysis and simulation}
One possible route to address this is to explore observational data of extremely metal poor (EMP) stars, 
defined by $[{\rm Fe/H}] < -3$. To this end, we consider objects from the SAGA\footnote{http://sagadatabase.jp/} 
database that are on the main sequence based on the effective temperature and the surface gravity. We find 
that there are a few stars that have characteristics corresponding to mass lower than $0.8\,{\rm M_{\odot}}$ 
on this plane. In order to interpret data on this plane, we examined results from MESA\footnote{http://mesa-web.asu.edu/} 
for metal poor stars showing that low-metallicity stars tend to have a higher effective temperature as 
compared to Population I stars of the same mass. The difference in effective temperature is about 
$30\%$, while the change in surface gravity is subdominant. Further, we verified that stars with mass 
less than $0.78\msol$ have an age on the main sequence that is comparable to the age of the 
universe (in preparation). Hence, our use of a mass threshold of $0.8\msol$ is close to the true value. 
Thus, there are few constraints on the initial mass function of metal poor stars at present. We expect that 
this will change with determination of distances to halo stars with GAIA.

\section*{Acknowledgements}

The authors thank the anonymous referee for providing a timely and constructive report that helped us 
to improve the quality of the paper. J.D. thanks Prof.~Biman Nath for helpful comments. J.D. would also like to 
acknowledge the Raman Research Institute and the Inter-University Center for Astronomy and Astrophysics 
for arranging visits and seminar for plenty of useful discussions, partial financial support, and local hospitality. 
The work of J.D. is supported by the Science and Engineering Research Board (SERB) of the Department of 
Science and Technology (DST), Government of India, as a National Post Doctoral Fellowship (NPDF) at the 
Physical Science department of IISER Mohali. We acknowledge the use of HPC facilities at IISER Mohali. This 
research has made use of NASA's Astrophysics Data System Bibliographic Services.


\begin{thebibliography}{}
\expandafter\ifx\csname natexlab\endcsname\relax\def\natexlab#1{#1}\fi
\providecommand{\url}[1]{\href{#1}{#1}}
\providecommand{\dodoi}[1]{doi:~\href{http://doi.org/#1}{\nolinkurl{#1}}}
\providecommand{\doeprint}[1]{\href{http://ascl.net/#1}{\nolinkurl{http://ascl.net/#1}}}
\providecommand{\doarXiv}[1]{\href{https://arxiv.org/abs/#1}{\nolinkurl{https://arxiv.org/abs/#1}}}

\bibitem[{{Aarseth} \& {Heggie}(1998)}]{aarseth98}
{Aarseth}, S.~J., \& {Heggie}, D.~C. 1998, \mnras, 297, 794,
  \dodoi{10.1046/j.1365-8711.1998.01521.x}

\bibitem[{{Abel} {et~al.}(2002){Abel}, {Bryan}, \& {Norman}}]{abn02}
{Abel}, T., {Bryan}, G.~L., \& {Norman}, M.~L. 2002, Science, 295, 93,
  \dodoi{10.1126/science.295.5552.93}

\bibitem[{{Bagla} {et~al.}(2009){Bagla}, {Kulkarni}, \& {Padmanabhan}}]{bkp09}
{Bagla}, J.~S., {Kulkarni}, G., \& {Padmanabhan}, T. 2009, \mnras, 397, 971,
  \dodoi{10.1111/j.1365-2966.2009.15012.x}

\bibitem[{{Barrow} {et~al.}(2017){Barrow}, {Wise}, {Norman}, {O'Shea}, \&
  {Xu}}]{barrow17}
{Barrow}, K. S.~S., {Wise}, J.~H., {Norman}, M.~L., {O'Shea}, B.~W., \& {Xu},
  H. 2017, \mnras, 469, 4863, \dodoi{10.1093/mnras/stx1181}

\bibitem[{{Becerra} {et~al.}(2015){Becerra}, {Greif}, {Springel}, \&
  {Hernquist}}]{bgsh15}
{Becerra}, F., {Greif}, T.~H., {Springel}, V., \& {Hernquist}, L.~E. 2015,
  \mnras, 446, 2380, \dodoi{10.1093/mnras/stu2284}

\bibitem[{{Bland-Hawthorn} {et~al.}(2015){Bland-Hawthorn}, {Sutherland}, \&
  {Webster}}]{bsw15}
{Bland-Hawthorn}, J., {Sutherland}, R., \& {Webster}, D. 2015, \apj, 807, 154,
  \dodoi{10.1088/0004-637X/807/2/154}

\bibitem[{{Bobrick} {et~al.}(2017){Bobrick}, {Davies}, \& {Church}}]{bdc17}
{Bobrick}, A., {Davies}, M.~B., \& {Church}, R.~P. 2017, \mnras, 467, 3556,
  \dodoi{10.1093/mnras/stx312}

\bibitem[{{Bond}(1981)}]{bond81}
{Bond}, H.~E. 1981, \apj, 248, 606, \dodoi{10.1086/159186}

\bibitem[{{Bondi}(1952)}]{bondi52}
{Bondi}, H. 1952, \mnras, 112, 195, \dodoi{10.1093/mnras/112.2.195}

\bibitem[{{Bondi} \& {Hoyle}(1944)}]{bondi44}
{Bondi}, H., \& {Hoyle}, F. 1944, \mnras, 104, 273,
  \dodoi{10.1093/mnras/104.5.273}

\bibitem[{{Bonifacio} {et~al.}(2018){Bonifacio}, {Caffau}, {Spite}, {Spite},
  {Sbordone}, {Monaco}, {Fran{\c{c}}ois}, {Plez}, {Molaro}, {Gallagher},
  {Cayrel}, {Christlieb}, {Klessen}, {Koch}, {Ludwig}, {Steffen}, {Zaggia}, \&
  {Abate}}]{bonifacio18}
{Bonifacio}, P., {Caffau}, E., {Spite}, M., {et~al.} 2018, \aap, 612, A65,
  \dodoi{10.1051/0004-6361/201732320}

\bibitem[{{Bromm} \& {Loeb}(2003)}]{bl03}
{Bromm}, V., \& {Loeb}, A. 2003, \nat, 425, 812, \dodoi{10.1038/nature02071}

\bibitem[{{Bromm} \& {Yoshida}(2011)}]{by11}
{Bromm}, V., \& {Yoshida}, N. 2011, \araa, 49, 373,
  \dodoi{10.1146/annurev-astro-081710-102608}

\bibitem[{{Brook} {et~al.}(2007){Brook}, {Kawata}, {Scannapieco}, {Martel}, \&
  {Gibson}}]{brook07}
{Brook}, C.~B., {Kawata}, D., {Scannapieco}, E., {Martel}, H., \& {Gibson},
  B.~K. 2007, \apj, 661, 10, \dodoi{10.1086/511514}

\bibitem[{{Caffau} {et~al.}(2011){Caffau}, {Bonifacio}, {Fran{\c{c}}ois},
  {Sbordone}, {Monaco}, {Spite}, {Spite}, {Ludwig}, {Cayrel}, {Zaggia},
  {Hammer}, {Randich}, {Molaro}, \& {Hill}}]{caffau11}
{Caffau}, E., {Bonifacio}, P., {Fran{\c{c}}ois}, P., {et~al.} 2011, \nat, 477,
  67, \dodoi{10.1038/nature10377}

\bibitem[{{Clark} {et~al.}(2011{\natexlab{a}}){Clark}, {Glover}, {Klessen}, \&
  {Bromm}}]{cgkb11}
{Clark}, P.~C., {Glover}, S. C.~O., {Klessen}, R.~S., \& {Bromm}, V.
  2011{\natexlab{a}}, \apj, 727, 110, \dodoi{10.1088/0004-637X/727/2/110}

\bibitem[{{Clark} {et~al.}(2011{\natexlab{b}}){Clark}, {Glover}, {Smith},
  {Greif}, {Klessen}, \& {Bromm}}]{cgsgkb11}
{Clark}, P.~C., {Glover}, S. C.~O., {Smith}, R.~J., {et~al.}
  2011{\natexlab{b}}, Science, 331, 1040, \dodoi{10.1126/science.1198027}

\bibitem[{{de Bennassuti} {et~al.}(2017){de Bennassuti}, {Salvadori},
  {Schneider}, {Valiante}, \& {Omukai}}]{deb17}
{de Bennassuti}, M., {Salvadori}, S., {Schneider}, R., {Valiante}, R., \&
  {Omukai}, K. 2017, \mnras, 465, 926, \dodoi{10.1093/mnras/stw2687}

\bibitem[{{de Souza} {et~al.}(2014){de Souza}, {Ishida}, {Whalen}, {Johnson},
  \& {Ferrara}}]{desouza14}
{de Souza}, R.~S., {Ishida}, E.~E.~O., {Whalen}, D.~J., {Johnson}, J.~L., \&
  {Ferrara}, A. 2014, \mnras, 442, 1640, \dodoi{10.1093/mnras/stu984}

\bibitem[{{Diemand} {et~al.}(2005){Diemand}, {Moore}, \& {Stadel}}]{dms05}
{Diemand}, J., {Moore}, B., \& {Stadel}, J. 2005, \nat, 433, 389,
  \dodoi{10.1038/nature03270}

\bibitem[{{Dutta}(2015)}]{dutta15}
{Dutta}, J. 2015, \apj, 811, 98, \dodoi{10.1088/0004-637X/811/2/98}

\bibitem[{{Dutta}(2016{\natexlab{a}})}]{dutta16a}
---. 2016{\natexlab{a}}, \aap, 585, A59, \dodoi{10.1051/0004-6361/201526747}

\bibitem[{{Dutta}(2016{\natexlab{b}})}]{dutta16b}
---. 2016{\natexlab{b}}, \apss, 361, 35, \dodoi{10.1007/s10509-015-2622-y}

\bibitem[{{Dutta} {et~al.}(2015){Dutta}, {Nath}, {Clark}, \&
  {Klessen}}]{dnck15}
{Dutta}, J., {Nath}, B.~B., {Clark}, P.~C., \& {Klessen}, R.~S. 2015, \mnras,
  450, 202, \dodoi{10.1093/mnras/stv664}

\bibitem[{{Edgar}(2004)}]{edgar04}
{Edgar}, R. 2004, \nar, 48, 843, \dodoi{10.1016/j.newar.2004.06.001}

\bibitem[{{Frebel} \& {Norris}(2015)}]{frebel15}
{Frebel}, A., \& {Norris}, J.~E. 2015, \araa, 53, 631,
  \dodoi{10.1146/annurev-astro-082214-122423}

\bibitem[{{Gao} {et~al.}(2007){Gao}, {Yoshida}, {Abel}, {Frenk}, {Jenkins}, \&
  {Springel}}]{gao07}
{Gao}, L., {Yoshida}, N., {Abel}, T., {et~al.} 2007, \mnras, 378, 449,
  \dodoi{10.1111/j.1365-2966.2007.11814.x}

\bibitem[{{Glover}(2015)}]{g15}
{Glover}, S. C.~O. 2015, \mnras, 451, 2082, \dodoi{10.1093/mnras/stv1059}

\bibitem[{{Greif}(2015)}]{greif15}
{Greif}, T.~H. 2015, Computational Astrophysics and Cosmology, 2, 3,
  \dodoi{10.1186/s40668-014-0006-2}

\bibitem[{{Greif} {et~al.}(2012){Greif}, {Bromm}, {Clark}, {Glover}, {Smith},
  {Klessen}, {Yoshida}, \& {Springel}}]{greif12}
{Greif}, T.~H., {Bromm}, V., {Clark}, P.~C., {et~al.} 2012, \mnras, 424, 399,
  \dodoi{10.1111/j.1365-2966.2012.21212.x}

\bibitem[{{Griffen} {et~al.}(2018){Griffen}, {Dooley}, {Ji}, {O'Shea},
  {G{\'o}mez}, \& {Frebel}}]{griffen18}
{Griffen}, B.~F., {Dooley}, G.~A., {Ji}, A.~P., {et~al.} 2018, \mnras, 474,
  443, \dodoi{10.1093/mnras/stx2749}

\bibitem[{{Haemmerl{\'e}} {et~al.}(2018){Haemmerl{\'e}}, {Woods}, {Klessen},
  {Heger}, \& {Whalen}}]{haemmerlE18}
{Haemmerl{\'e}}, L., {Woods}, T.~E., {Klessen}, R.~S., {Heger}, A., \&
  {Whalen}, D.~J. 2018, \mnras, 474, 2757, \dodoi{10.1093/mnras/stx2919}

\bibitem[{{Haiman}(2011)}]{haiman11}
{Haiman}, Z. 2011, \nat, 472, 47, \dodoi{10.1038/472047a}

\bibitem[{{Hartwig} {et~al.}(2019){Hartwig}, {Ishigaki}, {Klessen}, \&
  {Yoshida}}]{hartwig19}
{Hartwig}, T., {Ishigaki}, M.~N., {Klessen}, R.~S., \& {Yoshida}, N. 2019,
  \mnras, 482, 1204, \dodoi{10.1093/mnras/sty2783}

\bibitem[{{Hashimoto} {et~al.}(2018){Hashimoto}, {Laporte}, {Mawatari},
  {Ellis}, {Inoue}, {Zackrisson}, {Roberts-Borsani}, {Zheng}, {Tamura},
  {Bauer}, {Fletcher}, {Harikane}, {Hatsukade}, {Hayatsu}, {Matsuda}, {Matsuo},
  {Okamoto}, {Ouchi}, {Pell{\'o}}, {Rydberg}, {Shimizu}, {Taniguchi},
  {Umehata}, \& {Yoshida}}]{Hashimoto18}
{Hashimoto}, T., {Laporte}, N., {Mawatari}, K., {et~al.} 2018, \nat, 557, 392,
  \dodoi{10.1038/s41586-018-0117-z}

\bibitem[{{Hirano} \& {Bromm}(2017)}]{hb17}
{Hirano}, S., \& {Bromm}, V. 2017, \mnras, 470, 898,
  \dodoi{10.1093/mnras/stx1220}

\bibitem[{{Hirano} {et~al.}(2014){Hirano}, {Hosokawa}, {Yoshida}, {Umeda},
  {Omukai}, {Chiaki}, \& {Yorke}}]{hirano14}
{Hirano}, S., {Hosokawa}, T., {Yoshida}, N., {et~al.} 2014, \apj, 781, 60,
  \dodoi{10.1088/0004-637X/781/2/60}

\bibitem[{{Hosokawa} {et~al.}(2016){Hosokawa}, {Hirano}, {Kuiper}, {Yorke},
  {Omukai}, \& {Yoshida}}]{hosokawa16}
{Hosokawa}, T., {Hirano}, S., {Kuiper}, R., {et~al.} 2016, \apj, 824, 119,
  \dodoi{10.3847/0004-637X/824/2/119}

\bibitem[{{Ishiyama} {et~al.}(2016){Ishiyama}, {Sudo}, {Yokoi}, {Hasegawa},
  {Tominaga}, \& {Susa}}]{ishiyama16}
{Ishiyama}, T., {Sudo}, K., {Yokoi}, S., {et~al.} 2016, \apj, 826, 9,
  \dodoi{10.3847/0004-637X/826/1/9}

\bibitem[{{Johnson}(2015)}]{johnson15}
{Johnson}, J.~L. 2015, \mnras, 453, 2771, \dodoi{10.1093/mnras/stv1815}

\bibitem[{{Karlsson} {et~al.}(2013){Karlsson}, {Bromm}, \&
  {Bland-Hawthorn}}]{kbb13}
{Karlsson}, T., {Bromm}, V., \& {Bland-Hawthorn}, J. 2013, Reviews of Modern
  Physics, 85, 809, \dodoi{10.1103/RevModPhys.85.809}

\bibitem[{{Kirihara} {et~al.}(2019){Kirihara}, {Tanikawa}, \&
  {Ishiyama}}]{kti19}
{Kirihara}, T., {Tanikawa}, A., \& {Ishiyama}, T. 2019, \mnras, 486, 5917,
  \dodoi{10.1093/mnras/stz1277}

\bibitem[{{Komiya} {et~al.}(2009){Komiya}, {Habe}, {Suda}, \&
  {Fujimoto}}]{komiya09}
{Komiya}, Y., {Habe}, A., {Suda}, T., \& {Fujimoto}, M.~Y. 2009, \apjl, 696,
  L79, \dodoi{10.1088/0004-637X/696/1/L79}

\bibitem[{{Komiya} {et~al.}(2015){Komiya}, {Suda}, \& {Fujimoto}}]{komiya15}
{Komiya}, Y., {Suda}, T., \& {Fujimoto}, M.~Y. 2015, \apjl, 808, L47,
  \dodoi{10.1088/2041-8205/808/2/L47}

\bibitem[{{Komiya} {et~al.}(2016){Komiya}, {Suda}, \& {Fujimoto}}]{komiya16}
---. 2016, \apj, 820, 59, \dodoi{10.3847/0004-637X/820/1/59}

\bibitem[{{Krumholz} {et~al.}(2004){Krumholz}, {McKee}, \&
  {Klein}}]{krumholz04}
{Krumholz}, M.~R., {McKee}, C.~F., \& {Klein}, R.~I. 2004, \apj, 611, 399,
  \dodoi{10.1086/421935}

\bibitem[{{Latif} {et~al.}(2018){Latif}, {Volonteri}, \& {Wise}}]{latif18}
{Latif}, M.~A., {Volonteri}, M., \& {Wise}, J.~H. 2018, \mnras, 476, 5016,
  \dodoi{10.1093/mnras/sty622}

\bibitem[{{Machida} \& {Doi}(2013)}]{md13}
{Machida}, M.~N., \& {Doi}, K. 2013, \mnras, 435, 3283,
  \dodoi{10.1093/mnras/stt1524}

\bibitem[{{Magg} {et~al.}(2018){Magg}, {Hartwig}, {Agarwal}, {Frebel},
  {Glover}, {Griffen}, \& {Klessen}}]{magg18}
{Magg}, M., {Hartwig}, T., {Agarwal}, B., {et~al.} 2018, \mnras, 473, 5308,
  \dodoi{10.1093/mnras/stx2729}

\bibitem[{{Maki} \& {Susa}(2007)}]{maki07}
{Maki}, H., \& {Susa}, H. 2007, \pasj, 59, 787, \dodoi{10.1093/pasj/59.4.787}

\bibitem[{{Marigo} {et~al.}(2001){Marigo}, {Girardi}, {Chiosi}, \&
  {Wood}}]{marigo+01}
{Marigo}, P., {Girardi}, L., {Chiosi}, C., \& {Wood}, P.~R. 2001, \aap, 371,
  152, \dodoi{10.1051/0004-6361:20010309}

\bibitem[{{Monaghan}(1992)}]{monaghan92}
{Monaghan}, J.~J. 1992, \araa, 30, 543,
  \dodoi{10.1146/annurev.aa.30.090192.002551}

\bibitem[{{Nakamura} \& {Umemura}(2001)}]{nu01}
{Nakamura}, F., \& {Umemura}, M. 2001, \apj, 548, 19, \dodoi{10.1086/318663}

\bibitem[{{Omukai} \& {Nishi}(1998)}]{on98}
{Omukai}, K., \& {Nishi}, R. 1998, \apj, 508, 141, \dodoi{10.1086/306395}

\bibitem[{{O'Shea} \& {Norman}(2007)}]{on07}
{O'Shea}, B.~W., \& {Norman}, M.~L. 2007, \apj, 654, 66, \dodoi{10.1086/509250}

\bibitem[{{Salvadori} {et~al.}(2010){Salvadori}, {Ferrara}, {Schneider},
  {Scannapieco}, \& {Kawata}}]{salvadori10}
{Salvadori}, S., {Ferrara}, A., {Schneider}, R., {Scannapieco}, E., \&
  {Kawata}, D. 2010, \mnras, 401, L5, \dodoi{10.1111/j.1745-3933.2009.00772.x}

\bibitem[{{Scannapieco} {et~al.}(2006){Scannapieco}, {Kawata}, {Brook},
  {Schneider}, {Ferrara}, \& {Gibson}}]{scannapieco06}
{Scannapieco}, E., {Kawata}, D., {Brook}, C.~B., {et~al.} 2006, \apj, 653, 285,
  \dodoi{10.1086/508487}

\bibitem[{{Schlaufman} {et~al.}(2018){Schlaufman}, {Thompson}, \&
  {Casey}}]{schlaufman18}
{Schlaufman}, K.~C., {Thompson}, I.~B., \& {Casey}, A.~R. 2018, \apj, 867, 98,
  \dodoi{10.3847/1538-4357/aadd97}

\bibitem[{{Sharda} {et~al.}(2019){Sharda}, {Krumholz}, \&
  {Federrath}}]{sharda19}
{Sharda}, P., {Krumholz}, M.~R., \& {Federrath}, C. 2019, \mnras, 490, 513,
  \dodoi{10.1093/mnras/stz2618}

\bibitem[{{Sharma} {et~al.}(2018){Sharma}, {Theuns}, \& {Frenk}}]{stf18}
{Sharma}, M., {Theuns}, T., \& {Frenk}, C. 2018, \mnras, 477, L111,
  \dodoi{10.1093/mnrasl/sly052}

\bibitem[{{Shu}(1977)}]{shu77}
{Shu}, F.~H. 1977, \apj, 214, 488, \dodoi{10.1086/155274}

\bibitem[{{Sluder} {et~al.}(2016){Sluder}, {Ritter}, {Safranek-Shrader},
  {Milosavljevi{\'c}}, \& {Bromm}}]{sluder16}
{Sluder}, A., {Ritter}, J.~S., {Safranek-Shrader}, C., {Milosavljevi{\'c}}, M.,
  \& {Bromm}, V. 2016, \mnras, 456, 1410, \dodoi{10.1093/mnras/stv2587}

\bibitem[{{Sormani} {et~al.}(2017){Sormani}, {Tre{\ss}}, {Klessen}, \&
  {Glover}}]{sormani17}
{Sormani}, M.~C., {Tre{\ss}}, R.~G., {Klessen}, R.~S., \& {Glover}, S. C.~O.
  2017, \mnras, 466, 407, \dodoi{10.1093/mnras/stw3205}

\bibitem[{{Springel}(2010)}]{springel10sph}
{Springel}, V. 2010, \araa, 48, 391,
  \dodoi{10.1146/annurev-astro-081309-130914}

\bibitem[{{Springel} {et~al.}(2006){Springel}, {Frenk}, \&
  {White}}]{springel06}
{Springel}, V., {Frenk}, C.~S., \& {White}, S. D.~M. 2006, \nat, 440, 1137,
  \dodoi{10.1038/nature04805}

\bibitem[{{Stacy} {et~al.}(2016){Stacy}, {Bromm}, \& {Lee}}]{sbl16}
{Stacy}, A., {Bromm}, V., \& {Lee}, A.~T. 2016, \mnras, 462, 1307,
  \dodoi{10.1093/mnras/stw1728}

\bibitem[{{Stacy} {et~al.}(2013){Stacy}, {Greif}, {Klessen}, {Bromm}, \&
  {Loeb}}]{stacy13}
{Stacy}, A., {Greif}, T.~H., {Klessen}, R.~S., {Bromm}, V., \& {Loeb}, A. 2013,
  \mnras, 431, 1470, \dodoi{10.1093/mnras/stt264}

\bibitem[{{Sugimura} {et~al.}(2020){Sugimura}, {Matsumoto}, {Hosokawa},
  {Hirano}, \& {Omukai}}]{sugimura20}
{Sugimura}, K., {Matsumoto}, T., {Hosokawa}, T., {Hirano}, S., \& {Omukai}, K.
  2020, \apjl, 892, L14, \dodoi{10.3847/2041-8213/ab7d37}

\bibitem[{{Sur} {et~al.}(2010){Sur}, {Schleicher}, {Banerjee}, {Federrath}, \&
  {Klessen}}]{sur10}
{Sur}, S., {Schleicher}, D.~R.~G., {Banerjee}, R., {Federrath}, C., \&
  {Klessen}, R.~S. 2010, \apjl, 721, L134, \dodoi{10.1088/2041-8205/721/2/L134}

\bibitem[{{Susa}(2019)}]{susa19}
{Susa}, H. 2019, \apj, 877, 99, \dodoi{10.3847/1538-4357/ab1b6f}

\bibitem[{{Suto} \& {Silk}(1988)}]{suto88}
{Suto}, Y., \& {Silk}, J. 1988, \apj, 326, 527, \dodoi{10.1086/166114}

\bibitem[{{Syer} \& {White}(1998)}]{sw98}
{Syer}, D., \& {White}, S. D.~M. 1998, \mnras, 293, 337,
  \dodoi{10.1046/j.1365-8711.1998.01285.x}

\bibitem[{{Tanaka} {et~al.}(2017){Tanaka}, {Chiaki}, {Tominaga}, \&
  {Susa}}]{tanaka17}
{Tanaka}, S.~J., {Chiaki}, G., {Tominaga}, N., \& {Susa}, H. 2017, \apj, 844,
  137, \dodoi{10.3847/1538-4357/aa7e2c}

\bibitem[{{Tumlinson}(2010)}]{tumlinson10}
{Tumlinson}, J. 2010, \apj, 708, 1398, \dodoi{10.1088/0004-637X/708/2/1398}

\bibitem[{{Turk} {et~al.}(2009){Turk}, {Abel}, \& {O'Shea}}]{tao09}
{Turk}, M.~J., {Abel}, T., \& {O'Shea}, B. 2009, Science, 325, 601,
  \dodoi{10.1126/science.1173540}

\bibitem[{{Umeda} {et~al.}(2016){Umeda}, {Hosokawa}, {Omukai}, \&
  {Yoshida}}]{umeda16}
{Umeda}, H., {Hosokawa}, T., {Omukai}, K., \& {Yoshida}, N. 2016, \apjl, 830,
  L34, \dodoi{10.3847/2041-8205/830/2/L34}

\bibitem[{{Whalen} {et~al.}(2014){Whalen}, {Smidt}, {Even}, {Woosley}, {Heger},
  {Stiavelli}, \& {Fryer}}]{whalen14}
{Whalen}, D.~J., {Smidt}, J., {Even}, W., {et~al.} 2014, \apj, 781, 106,
  \dodoi{10.1088/0004-637X/781/2/106}

\bibitem[{{Wise}(2019)}]{wise19}
{Wise}, J.~H. 2019, arXiv e-prints, arXiv:1907.06653.
\newblock \doarXiv{1907.06653}

\bibitem[{{Wollenberg} {et~al.}(2020){Wollenberg}, {Glover}, {Clark}, \&
  {Klessen}}]{wollenberg20}
{Wollenberg}, K. M.~J., {Glover}, S. C.~O., {Clark}, P.~C., \& {Klessen}, R.~S.
  2020, \mnras, 494, 1871, \dodoi{10.1093/mnras/staa289}
  
\bibitem[{{Woods} {et~al.}(2017){Woods}, {Heger}, {Whalen}, {Haemmerl{\'e}}, \&
  {Klessen}}]{woods17}
{Woods}, T.~E., {Heger}, A., {Whalen}, D.~J., {Haemmerl{\'e}}, L., \&
  {Klessen}, R.~S. 2017, \apjl, 842, L6, \dodoi{10.3847/2041-8213/aa7412}

\bibitem[{{Xu} {et~al.}(2016){Xu}, {Ahn}, {Norman}, {Wise}, \& {O'Shea}}]{xu16}
{Xu}, H., {Ahn}, K., {Norman}, M.~L., {Wise}, J.~H., \& {O'Shea}, B.~W. 2016,
  \apjl, 832, L5, \dodoi{10.3847/2041-8205/832/1/L5}

\bibitem[{{Yoshida} {et~al.}(2008){Yoshida}, {Omukai}, \& {Hernquist}}]{yoh08}
{Yoshida}, N., {Omukai}, K., \& {Hernquist}, L. 2008, Science, 321, 669,
  \dodoi{10.1126/science.1160259}

\bibitem[{{Yoshida} {et~al.}(2006){Yoshida}, {Omukai}, {Hernquist}, \&
  {Abel}}]{yoha06}
{Yoshida}, N., {Omukai}, K., {Hernquist}, L., \& {Abel}, T. 2006, \apj, 652, 6,
  \dodoi{10.1086/507978}

\end{thebibliography}
\end{document}